\documentclass[11pt,a4paper]{article}
\pdfoutput=1
\usepackage{jheppub}
\usepackage{amsfonts}
\usepackage{amssymb}
\usepackage{amscd}

\title{Black Hole Thermodynamics Based on Unitary Evolutions}

\author[a]{Yu-Lei Feng,}
\author[a,1]{Yi-Xin Chen,\note{Corresponding author.}}

\affiliation[a]{Zhejiang Institute of Modern Physics, Zhejiang University,\\
Hangzhou, 310027, P. R. China}
\emailAdd{11336008@zju.edu.cn}
\emailAdd{yixinchenzimp@zju.edu.cn}

\abstract{ In this paper, we try to construct black hole thermodynamics based on the fact that, the formation and evaporation of a black hole can be described by quantum unitary evolutions. First, we show that the Bekenstein-Hawking entropy $S_{BH}$ may not be a Boltzmann or thermal entropy. To confirm this statement, we show that the original black hole's ``first law" may not simply be treated as the first law of thermodynamics formally, due to some missing metric perturbations caused by matter. Then, by including those (quantum) metric perturbations, we show that the black hole formation and evaporation can be described in a unitary manner effectively, through a quantum channel between the exterior and interior of the event horizon. In this way, the paradoxes of information loss and firewall can be resolved effectively. Finally, we show that black hole thermodynamics can be constructed in an ordinary way, by constructing statistical mechanics.
}
\begin{document}
\maketitle
\flushbottom

\section{Introduction}
\label{seci}
Black hole is a mysterious object, in particular when Hawking found that it can evaporate and leaves thermal radiation~\cite{a,b}. After that, black hole thermodynamics~\cite{c} becomes more attractive, which is regarded to be a clue of the unknown quantum gravity theory. Among those thermodynamic quantities, the black hole entropy seems to be more important and its meaning is still unclear. The black hole entropy is usually expressed as a Bekenstein-Hawking form~\cite{d}
\begin{equation}
S_{BH}=A/4\hbar G,
\label{a}
\end{equation}
with $A$ the area of the black hole's event horizon. Besides, there is a corresponding Hawking temperature~\cite{a}
\begin{equation}
T_{H}=\hbar\kappa/2\pi,
\label{b}
\end{equation}
with $\kappa$ the surface gravity on the event horizon.

The derivations of $S_{BH}$ are roughly divided into two classes. One class is based on the classical Einstein's equation, which is also the original derivation of ``black hole thermodynamics"~\cite{c}. In particular, the first law takes the form
\begin{equation}
\delta M=\kappa \delta A/8\pi G+\Omega \delta J+\Phi \delta Q,
\label{b1}
\end{equation}
for a general rotating charged black hole, from which the roles of entropy and temperature for $S_{BH}$ and $T_{H}$ can be derived formally, by comparing with the first law of thermodynamics. The other class is based on the quantum von Neumann entropy or entanglement entropy~\cite{e,f}
\begin{equation}
S_{en}=-Tr(\hat{\rho}_{ext}\ln\hat{\rho}_{ext}),
\label{c}
\end{equation}
with $\hat{\rho}_{ext}$ a reduced density matrix for the exterior field modes, after partially tracing over the unobserved interior modes within the in-falling vacuum state~\cite{g,h,i,i1,j,j1}. There are also some other attempts based on counting the underlying microscopic degrees of freedom, for example superstring theory~\cite{k} and holographic hypothesis~\cite{l,m}. The reason for this classification is that the first class tends to treat $S_{BH}$ as a Boltzmann or thermal entropy, counting microstates according to statistical mechanics. While the second class uses the concepts in quantum mechanics and quantum information theory.

Although the entanglement entropy in Eq.~(\ref{c}) gives the area law accurately up to some corrections, there are still some problems in the interpretation of Bekenstein-Hawking entropy as entanglement entropy. For example~\cite{j1}, entanglement entropy is a UV divergent quantity, while the Bekenstein-Hawking entropy is a finite quantity; the entanglement entropy is proportional to the number of field species, but the Bekenstein-Hawking entropy doesn't seem to depend on any number of fields, etc. Thus, it seems that $S_{BH}$ is a Boltzmann or thermal entropy of the black hole.
However, if $S_{BH}$ is a Boltzmann entropy together with $T_{H}$ as a thermal temperature, the information loss paradox~\cite{a,b} can not be easily resolved. Actually, the ``black hole thermodynamics" in reference~\cite{c} is based on the ``non-unitary" Hawking evaporation that gives thermal radiations.

It seems that entanglement plays a crucial role in the black hole evaporation. In our recent paper~\cite{n}, a unitary model for black hole evaporation is proposed to transfer information between the black hole's interior and exterior, through the entanglement implicit in the in-falling vacuum state, i.e. a (modified) quantum teleportation\footnote{The model proposed in reference~\cite{n} can also resolve the paradox of firewall~\cite{n1,n2} or energetic curtains~\cite{n3}, see also the discussions in Sec.~\ref{seci1-2}.}.
We can also consider the problem just according to the principles of quantum mechanics. It is believed that \emph{all evolutions should be unitary for a closed system}. By treating the gravity-matter system as a closed system, the black hole formation and evaporation must be unitary. This abstract picture should be exact in a quantum gravity theory, but not in the framework of effective field theory with dependent background. This is because, in the latter case, classical black hole background is utilized so that its event horizon forms a causal barrier between its interior and exterior. But this barrier can be tunnelled effectively through quantum teleportation via entanglement, so that energy or information can be transferred between the two sides of the event horizon~\cite{n}. Therefore, the black hole thermodynamics may be constructed according to the unitary evolutions of the black hole-matter system, indicating that $S_{BH}$ is not a Boltzmann or thermal entropy, but just an entanglement entropy.

In this paper, we try to construct black hole thermodynamics based on unitary evolutions. First, we reconsider the original derivation of the ``first law" in Eq.~(\ref{b1}) in reference~\cite{c}. We find that the ``first law" may not be treated as a thermal one, due to some missing metric perturbations caused by matter. Actually, after including these metric perturbations, the total mass formula for the black hole-matter system may also contain a deformation term $\kappa \delta A/4\pi G$ in some particular case. This means that there will be a quadratic variation $\kappa \delta^2 A/4\pi G$ added into the total mass variation, so that $S_{BH}$ and $T_{H}$ may not be black hole's thermal quantities by only comparing with the first law of thermodynamics formally. Then, by including the (quantum) metric perturbations, we find that the black hole formation and evaporation can be described in a unitary manner effectively. This is achieved through a quantum channel between the black hole's exterior and interior, so that energy or information can be transported in and out more freely. In this way, the paradoxes of information loss and firewall can also be resolved effectively. Finally, we show that black hole thermodynamics can be derived ordinarily, by constructing the system's statistical mechanics\footnote{Some general investigations about the black hole statistical mechanics are also given in reference~\cite{n3a}. But the arguments there are not based on satisfactory unitary evolutions.}.

This paper is organized as follows. In Sec.~\ref{seci1-0}, we reconsider the original derivation of the black hole's ``first law", and point out the missing metric perturbations caused by matter. With the metric perturbations included, $S_{BH}$ and $T_{H}$ may no longer be treated as thermal quantities of the black hole. In Sec.~\ref{seci1-1} and Sec.~\ref{seci1-2}, we show that the event horizon can be tunneled through a quantum teleportation channel, so that energy can be transported between the two sides of the event horizon more efficiently. Then in Sec.~\ref{seci1-3}, based on the quantum tunneling, the black hole thermodynamics is constructed formally by constructing the black hole-matter system's statistical mechanics.  Finally, in Sec.~\ref{secv}, we summarize our results and draw some conclusions.

\section{Bekenstein-Hawking entropy may not be Boltzmann entropy}
\label{seci1-0}

When the Bekenstein-Hawking Entropy $S_{BH}$ is treated as a Boltzmann or thermal entropy together with the Hawking temperature $T_{H}$ as a thermal temperature, the resulted thermodynamics seems not to be the familiar one. Actually, when there is an environment containing matter, for example a heat bath, the thermal equilibrium between black hole and matter is unstable~\cite{n4}. A small fluctuation to larger mass will cause its temperature to drop, which leads to a runaway growth of the hole. Conversely, a small fluctuation to smaller mass will lead to a runaway evaporation of the hole. Certainly, this instability may be controlled in some way artificially, such as by putting the black hole in a very small container, and somehow holding the temperature at the box fixed. Easily to see, the instability is mainly due to the fact that the Hawking temperature $T_{H}$ depends on the black hole's mass. This indicates that the black hole's ``first law" in Eq.~(\ref{b1}) may not be a thermal one. And to see this, let's recall the original derivation in reference~\cite{c}.

For simplicity, we consider a Schwarzschild black hole with a mass $M_1$, together with some matter \emph{only outside} the event horizon. The derivation is based on the equation for a time translational Killing vector $K^\mu$
\begin{equation}
K^{\mu;\nu}~_{\nu}=-R^{\mu}~_{\nu}K^\nu,
\label{c1}
\end{equation}
where a semicolon denotes the covariant derivatives. Since $K_{\mu;\nu}$ is antisymmetric, one can integrate the above equation over a hypersurface $S$ and transfer the volume integral on the left to an integral over a 2-surface $\partial S$ bounding $S$
\begin{equation}
\int_{\partial S} K^{\mu;\nu}d\Sigma_{\mu\nu}=-\int_S R^{\mu}~_{\nu}K^\nu d\Sigma_{\mu},
\label{c2}
\end{equation}
where $d\Sigma_{\mu\nu}$ and $d\Sigma_{\mu}$ are the surface elements of $\partial S$ and $S$ respectively. The boundary $\partial S$ of $S$ consists of $\partial B$, the intersection of $S$ at the event horizon, and a 2-surface $\partial S_\infty$ at infinity. Then we will obtain
\begin{equation}
M=\int_S (2T^{\mu}~_{\nu}-T\delta^{\mu}~_{\nu})K^\nu d\Sigma_{\mu}+\frac{1}{4\pi G}\int_{\partial B} K^{\mu;\nu}d\Sigma_{\mu\nu},
\label{c3}
\end{equation}
where $M$ is the total mass as measured from infinity, and the Einstein's equation has been used. The first integral on the right can be regarded as the contribution to the
total mass of the matter outside the event horizon, and the second integral may be regarded as the mass $M_1$ of the black hole. One can further express $d\Sigma_{\mu\nu}$ as $K_{[\mu}n_{\nu]}dA$, where $n_\mu$ is the other null vector orthogonal to $\partial B$, normalized so that $n_\mu K^\mu=-1$, and $dA$ is the surface area element of $\partial B$. Thus the
last term on the right of Eq.~(\ref{c3}) is
\begin{equation}
\frac{1}{4\pi G}\int_{\partial B} \kappa_1 dA=\frac{1}{4\pi G} \kappa_1 A_1=M_1,
\label{c4}
\end{equation}
where $\kappa_1(=-K_{\mu;\nu}n^\mu K^\nu)$ is the constant surface gravity of the black hole at the event horizon. The first law for this black hole together with the matter can be derived further by varying Eq.~(\ref{c3}) between two slightly different stationary black hole solutions~\cite{c}.

Note that the factorized form of the black hole mass formula $M_1=\kappa_1 A_1/4\pi G$ is only proper for a \emph{pure} black hole without any matter. This can simply be verified for a pure (Schwarzschild) black hole with $\kappa_1=1/4GM_1$ and $A_1=4\pi(2GM_1)^2$. However, when there is some matter outside the event horizon, the general expression $\int_{\partial B} K^{\mu;\nu}d\Sigma_{\mu\nu}/4\pi G$ in Eq.~(\ref{c3}) will \emph{no longer} be expressed simply as the factorized form $\kappa_1 A_1/4\pi G$. This can be seen by noting that $R_{\mu\nu}\neq0$ outside the event horizon under this circumstance, due to the matter outside.
In other words, if there is some matter outside the event horizon, they will cause some perturbation to the (Schwarzschild) black hole's metric $g^B_{\mu\nu}$ with $R_{\mu\nu}[g^B]=0$, so that the final metric is
\begin{equation}
g_{\mu\nu}=g^B_{\mu\nu}+h^m_{\mu\nu},
\label{d1}
\end{equation}
where $h^m_{\mu\nu}$ is the contribution from matter. Then, the curvature in Eq.~(\ref{c1}) is $R_{\mu\nu}[g]$, similarly for the matter term $T_{\mu\nu}[g]$ and the Killing vector $K^\mu[g]$. Moreover, the event horizon for $g^B_{\mu\nu}$, in particular the 2-surface $\partial B$, will be deformed. As a result, $-K_{\mu;\nu}n^\mu K^\nu$ will not be just a constant surface gravity of the (pure) black hole. That is, the factorized form $\kappa_1 A_1/4\pi G$ is not enough, when there is some matter outside the event horizon.

The deformation depends on the matter term or the explicit expression of $h^m_{\mu\nu}$.
When the deformation is small enough, we can expand the last term in Eq.~(\ref{c3}) formally as
\begin{equation}
\frac{1}{4\pi G}\int_{\partial B} K^{\mu;\nu}d\Sigma_{\mu\nu}=\frac{1}{4\pi G}\kappa_1 A_1+\int_{\partial B}dA~\mathcal{V}_{int}(g^B,h^m,K),
\label{d2}
\end{equation}
where the second term is the interaction between $g^B_{\mu\nu}$ and $h^m_{\mu\nu}$ at the event horizon, or the reaction of matter on the background. The interaction term can be further expressed as
\begin{equation}
\frac{1}{4\pi G}\int_{\partial B}\delta\kappa_1dA,\qquad \delta\kappa_1=-(K_{\mu;\nu}n^\mu K^\nu)[g]+(K_{\mu;\nu}n^\mu K^\nu)[g^B],
\label{d2'}
\end{equation}
where $\delta\kappa_1$ is the density of the surface gravity deformation at the event horizon.
By substituting the metric in Eq.~(\ref{d1}), the matter contribution term in Eq.~(\ref{c3}) can be expressed as
\begin{equation}
\int_S d\Sigma_{\mu}(T^{\mu}_{\nu}-\frac{1}{2}T\delta^{\mu}_{\nu})K^\nu[g^B]+\frac{1}{8\pi G}\int_S dV~\mathcal{K}(h^m,K)[g^B]+\int_S dV~\mathcal{U}_{int}(h^m,T,K)[g^B],
\label{d3}
\end{equation}
i.e. the matter mass, the energy of the gravitational field $h^m_{\mu\nu}$, and the interaction between them at the background of the black hole\footnote{Note that there is a factor 2 in the matter term of Eq.~(\ref{c3}). And the energy of the gravitational field $h^m_{\mu\nu}$ comes from the expansion $R_{\mu\nu}[g^B+h^m]$, together with the fact $R_{\mu\nu}[g^B]=0$. }. Ultimately, the mass formula in Eq.~(\ref{c3}) will be rewritten formally as
\begin{equation}
M=M^B_m+M_1+\frac{1}{4\pi G}\int_{\partial B}\delta\kappa_1dA+E^B_h+U^B_{int},
\label{d4}
\end{equation}
where $E^B_h$ and $U^B_{int}$ are denoted as the energies of the metric perturbation (or gravitational field) $h^m_{\mu\nu}$ and the interaction in Eq.~(\ref{d3}) respectively.
Now, if there is some variation on the matter distribution outside the black hole, then $h^m_{\mu\nu}$ will also be varied. There are roughly two cases. If no matter falls into black hole, then we have
\begin{equation}
\delta M=\delta M^B_m+\frac{1}{4\pi G}\int_{\partial B}\delta^2\kappa_1dA+\delta E^B_h+\delta U^B_{int}.
\label{e1}
\end{equation}
If there is some matter falling into the black hole, leading to another black hole with a mass $M_2=\kappa_2 A_2/4\pi G$, then we have
\begin{equation}
\delta M=\delta M^{B}_m+(M_2-M_1)+\frac{1}{4\pi G}(\int_{\partial B_2}\delta\kappa_2dA-\int_{\partial B_1}\delta\kappa_1dA)+\delta E^B_h+\delta U^B_{int},
\label{e2}
\end{equation}
where we have used $\partial B_2$ and $\partial B_1$ to denote the two event horizons for the respective black holes.

When $M_2-M_1$ is small enough, it can be identified with $\kappa \delta A/8\pi G$ up to first order. In reference~\cite{c}, $\kappa \delta A/8\pi G$ is treated as $T_H\delta S_{BH}$ by comparing with the first law of thermodynamics formally. Then the formula of total mass variation is treated as the ``first law" for the black hole-matter system. In our derivation, however, by including the metric perturbations caused by matter, there is a deformation term $\int_{\partial B}\delta\kappa_1dA/4\pi G$ in the total mass formula in Eq.~(\ref{d4}). To see what can be brought about by this term, let's see a particular case, i.e. when the deformation is \emph{homogeneous} over the event horizon. Then we will have $\delta\kappa_1A_1/4\pi G=-\kappa_1\delta A_1/4\pi G$, since $\delta(\kappa_1A_1/4\pi G)=\delta M_1=0$. This implies further that some quadratic variation $\delta^2 A$ will occur in the total mass variation in Eq.~(\ref{e1}), and some contribution $-(\kappa_2 \delta A_2-\kappa_1 \delta A_1)/4\pi G$ will occur in Eq.~(\ref{e2}). This indicates that the area variation $\delta A$ here can no longer simply be treated as the variation of an entropy. That is, neither Eq.~(\ref{e1}) nor Eq.~(\ref{e2}) can simply be treated as the first law of thermodynamics formally. Moreover, the surface gravity deformation also indicates that the relationship $\kappa\sim T_H$ may not be appropriate. As a result, \emph{the Bekenstein-Hawking Entropy $S_{BH}$, which is roughly identified with the event horizon's area $A$, may not be treated as a Boltzmann or thermal entropy for a (Schwarzschild) black hole}. This implies further that the first law of thermodynamic for a black hole may not simply be derived by only comparing the mass variation with the thermal first law. In Sec.~\ref{seci1-3}, the black hole thermodynamics will be obtained by constructing the system's statistical mechanics.

\section{Unitary evolutions of the black hole-matter system}
\label{seci1}

It seems that the non-unitary of Hawking evaporation is mainly resulted from the causal structure of the black hole. Thus, to obtain unitary evolutions, the black hole's event horizon should be tunnelled effectively. In reference~\cite{n}, a model was proposed so that the event horizon barrier could be tunnelled effectively, by means of the entanglements implicit in the in-falling vacuum. In this section, we show the (quantum) tunneling according to some more general investigations. Then based on this tunneling, the black hole formation and evaporation can be described in a unitary manner effectively.

\subsection{Tunneling over the event horizon barrier: classical case}
\label{seci1-1}

Consider an intermediate stage in forming a larger black hole with a mass $M_2$ from a smaller one with a mass $M_1$. Suppose that the in-falling matter has crossed the event horizon but still far away from the singularity $r=0$. In the view of an exterior static observer, the total mass will be
\begin{equation}
M'=[M'^B_m]^{ext}+M_1+\frac{1}{4\pi G}\int_{\partial B}\delta'\kappa_1dA+[E'^B_h]^{ext}+[U'^B_{int}]^{ext}.
\label{f1}
\end{equation}
Note that in this intermediate stage, the black hole's mass is still $M_1$ since the in-falling matter is still far away from the singularity $r=0$\footnote{Notice that $R_{\mu\nu}\neq0$ also inside the event horizon, due to the in-falling matter. That is, the final (pure) black hole with mass $M_2$ has not been formed.}. Because of the singularity of the event horizon at $r=2M_1$, the other observed quantities on the right hand side of Eq.~(\ref{f1}) contain only exterior effects, except the deformation term $\int_{\partial B}\delta'\kappa_1dA/4\pi G$\footnote{Note that the interior effects cannot propagate outside the event horizon classically, due to the black hole's causal structure.}. By comparing with Eq.~(\ref{d4}), one can see that this deformation actually represents the total effects of both the exterior and interior matter, since some matter has fallen into the black hole. In fact, in this intermediate stage, the metric will be roughly deformed to be
\begin{equation}
g_{\mu\nu}=g^B_{\mu\nu}+[h^m_{\mu\nu}]^{ext}+[h^m_{\mu\nu}]^{int},
\label{f2}
\end{equation}
including both the exterior and interior deformations or perturbations. If the event horizon was \emph{not} singular, then the deformation term $\int_{\partial B}\delta'\kappa_1dA/4\pi G$ could simply be divided into two parts $[\int_{\partial B}\delta'\kappa_1dA/4\pi G]^{ext}+[\int_{\partial B}\delta'\kappa_1dA/4\pi G]^{int}$, resulted from the two metric perturbations in Eq.~(\ref{f2}) respectively. Besides, the exterior and interior matter could interact with each other through the propagations of those metric perturbations, for example $\int dV[T^{\mu\nu}]^{ext}[h^m_{\mu\nu}]^{int}$ formally, similarly for those exterior and interior metric perturbations themselves, i.e. $\int dV[h^{\mu\nu}]^{ext}[h_{\mu\nu}]^{int}$ formally. In the real case, however, the event horizon is singular so that the two metric perturbations are confined within their own regions (classically), except the common event horizon where the interactions occur. This implies that the deformation term $\int_{\partial B}\delta'\kappa_1dA/4\pi G$ should also contain the effects of those interactions forbidden by the black hole's causal structure.

Now inside the event horizon, we can make an analogous procedure as the one in Eq.~(\ref{c1}) and Eq.~(\ref{c2}), by integrating over another hypersurface $S'$ whose boundaries are $\partial B$ and $\partial S_0$, where $\partial S_0$ is a 2-surface near the singularity $r=0$. Then after some similar derivations as those in the last section, we have
\begin{equation}
M_1+\frac{1}{4\pi G}\int_{\partial B}\delta'\kappa_1dA=[M'^B_m]^{int}+[E'^B_h]^{int}+[U'^B_{int}]^{int}+M_1,
\label{f3}
\end{equation}
where the $M_1$ on the left hand side is from the integral over $\partial B$, while the one on the right hand side is from the integral over $\partial S_0$. The deformation of the 2-surface $\partial S_0$ is neglected because we have assumed that the in-falling matter is still far away from the singularity $r=0$. Substituting Eq.~(\ref{f3}) into Eq.~(\ref{f1}), the total mass becomes
\begin{equation}
M'=[M'^B_m]^{ext}+[E'^B_h]^{ext}+[U'^B_{int}]^{ext}+M_1+[M'^B_m]^{int}+[E'^B_h]^{int}+[U'^B_{int}]^{int},
\label{f4}
\end{equation}
which is symmetric between the exterior and interior terms\footnote{Certainly, all the exterior matter can fall into the black hole. However, this is possible only classically, not the case in the quantum version due to quantum fluctuations, as shown in the next subsection.}. The expression in Eq.~(\ref{f4}) seems to be obtained directly by integrating over a larger hypersurface that spans both the exterior and interior of the black hole, up to the neighborhood of the singularity $r=0$\footnote{It should be stressed that this statement is not exact, otherwise there may be some divergence problems at the event horizon. Notice that Eq.~(\ref{f4}) is actually obtained by substituting Eq.~(\ref{f3}) into Eq.~(\ref{f1}), in this way the divergence problems at the event horizon have been avoided.}. That is, the boundaries are the $\partial S_\infty$ at infinity and the $\partial S_0$ near the singularity $r=0$. In this sense, the singular event horizon seems to disappear, or the event horizon is ``tunneled"\footnote{This fact was expressed in reference~\cite{n} by an extended postulate (ii) of the black hole complementarity (BHC): ``both of the exterior and interior regions of the black hole can be well described by
QFT in curved space, with the singularity $r=0$ excluded". This is such a crucial extension that a quantum channel can be constructed between two sides of the event horizon, as shown in the next subsection.}. Certainly, this ``tunnel" is only classical, based only on the fact that matter can fall into the black hole freely\footnote{The terminology (classical) ``tunnel" is used here to correspond to the quantum tunnel in the next subsection, since the event horizon can be treated as a (classical) causal barrier. Certainly, this classical ``tunnel" is possible due to the existence of the in-falling frame, so that the in-falling matter can cross the event horizon freely classically.}. A quantum tunnel will be shown in the next subsection, so that matter or energy can fall into and come out freely via a quantum channel.

Since $M_2-M_1$ can be expressed as the deformation of the 2-surface $\partial B$, thus the black hole with a mass $M_2$ can be formed provided that the total deformation $\int_{\partial B}\delta'\kappa_1dA/4\pi G$ in Eq.~(\ref{f1}) or Eq.~(\ref{f3}) exceeds $M_2-M_1$, in particular in the homogeneous case. This
usually happens when the in-falling matter is near the singularity $r=0$. In this case, the right hand side of Eq.~(\ref{f3}) is $M_2$ together with some residual tiny perturbations, such as quantum fluctuations. Note further that the total deformation $\int_{\partial B}\delta'\kappa_1dA/4\pi G$ results from the metric deformations or perturbations in Eq.~(\ref{f2}). It thus implies that one can construct new black hole metric from those perturbations effectively as follows\footnote{Eq.~(\ref{f5}) can be verified roughly by calculating $g^B_{\mu\nu}(M_2)-g^B_{\mu\nu}(M_1)$ when $M_2-M_1$ is small enough. Certainly, it's not an exact description of the process, which can only be given by a quantum gravity theory.}
\begin{equation}
g^B_{\mu\nu}(M_2)+h'_{\mu\nu}=g^B_{\mu\nu}(M_1)+[h^m_{\mu\nu}]^{ext}+[h^m_{\mu\nu}]^{int},
\label{f5}
\end{equation}
where $h'_{\mu\nu}$ is some residual perturbation that cannot be used to form black holes with larger masses than $M_2$\footnote{The residual perturbation $h'_{\mu\nu}$ is necessary, in particular in the quantum case, since there are always some quantum fluctuations.}. Certainly, those interior perturbations are almost near the singularity $r=0$, where large perturbations can be generated by those fallen matter. Conversely, the black hole metric with larger mass, together with some small perturbations, can also be represented by another metric with smaller mass together with larger perturbations. This fact is important, because it provides a possibility for a black hole to lose its energy, leading to smaller black holes. Furthermore, since metric perturbations almost result from the matter distributions, it's necessary to consider the transport of matter or energy between black hole exterior and interior. Classically, only matter in-falling is allowed in an in-falling frame, leading to the classical tunnel in Eq.~(\ref{f4}), while matter can never escape from the event horizon interior. The Hawking effects may be a complement. Although the Hawking radiation's spectrum is thermal, it still indicates that, in the quantum case matter or energy may go into and come out more easily. This will be shown in the next subsection, where the paradoxes of information loss and firewall can also be resolved effectively.

\subsection{Tunneling over the event horizon barrier: quantum channel}
\label{seci1-2}

In the last subsection, we have shown that the event horizon can be ``tunneled" effectively, giving an expression in Eq.~(\ref{f4}) that is symmetric between the black hole exterior and interior. However, in the classical case matter can only be allowed to fall into the black hole interior, from which no signal can escape\footnote{Strictly speaking, in the static frame matter can not even fall into the black hole interior, in the sense of using infinite time. }. This asymmetry may be improved by Hawking effects. But the information loss paradox overshadows this possibility. In this subsection, we show that a unitary quantum channel can be constructed between the two sides of the event horizon, so that matter or energy can be transported in and out effectively .

To construct such a quantum channel, we should deal with some quantum fields. For simplicity, we consider only a massless scalar quantum matter field $\hat{\phi}$. Correspondingly, the metric perturbations due to the matter distributions should also be quantized, i.e. $\hat{h}_{\mu\nu}$. The mode expansions of these quantized fields at the black hole background are the familiar ones. For example, for the scalar field we have~\cite{s}
\begin{equation}
\hat{\phi} =  \int_{0}^{\infty}
d\omega(\hat{a}_{\omega}^{\dag}U_{\omega}^*+h.c.) = \int_{0}^{\infty}
d\omega(\hat{b}_{\omega}^{\dag}u_{\omega}^*+\hat{\tilde{b}}_{\omega}\tilde{u}_{\omega}+h.c.),
\label{g1}
\end{equation}
where $\hat{a}$ stands for the modes of the in-falling frame, while $\hat{b}$ and $\hat{\tilde{b}}$ are the exterior and interior modes respectively in the static frame. Expressions for the mode functions $U_{\omega}$, $u_{\omega}$, $\tilde{u}_{\omega}$ are not shown for simplicity. The in-falling vacuum of the scalar field is given by~\cite{s}
\begin{equation}
\arrowvert0_{U}\rangle_{\phi} = \prod_\omega(1-e^{-8\pi
M\omega})^{1/2}\exp(\sum_{\omega}e^{-4\pi
M\omega}\hat{b}_{\omega}^{\dag}\hat{\tilde{b}}_{\omega}^{\dag})\arrowvert0,\tilde{0}\rangle_{\phi},
\label{g2}
\end{equation}
at the background of a black hole with some mass $M$. The mode expansion for the metric perturbation $\hat{h}_{\mu\nu}$ is analogous, except for an added polarization $\epsilon_{\mu\nu}$. We use $\hat{c}$ to stands for the metric perturbation's modes in the in-falling frame, and $\hat{d}$ and $\hat{\tilde{d}}$ for its corresponding exterior and interior modes respectively in the static frame. Besides, we use subscript $h$ to denote the metric perturbation's states, for example, the in-falling vacuum $\arrowvert0_{U}\rangle_{h}$.

Given a black hole with a mass $M$, if the state of the metric perturbation $\hat{h}_{\mu\nu}$ is the in-falling vacuum $\arrowvert0_{U}\rangle_{h}$\footnote{It should be stressed that the entanglements implicit in those in-falling vacua cannot be destroyed, otherwise something like firewall would occur so that the space-time would be broken or ended along the event horizon.}, then the expectation value $_{h}\langle0_{U}\arrowvert\hat{h}_{\mu\nu}\arrowvert0_{U}\rangle_{h}$ is vanishing, and the metric is not deformed in this case. However, the expectation values of some quantum polynomials $\hat{P}[\hat{h}_{\mu\nu}]$, such as the Hamiltonian of $\hat{h}_{\mu\nu}$, will be non-vanishing so that the metric may be deformed. This is analogous to the Hawking effects with some non-vanishing $_{\phi}\langle0_{U}\arrowvert\hat{T}_{\mu\nu}[\hat{b}]\arrowvert0_{U}\rangle_{\phi}$ in the exterior region. However, for large enough black hole's mass $M$, these effects are only of order $M^{-2}$ so that their resulting deformations are not large enough to change the black hole greatly\footnote{When the mass $M$ is small, such as for an atomic black hole, the Hawking effects seem to be influential. However, in that case there would be a cutoff in the frequency integral, because no high energy particles can be created for such a light black hole. Certainly, only a complete quantum gravity theory can apply in this extreme case.}. In other words, the Hawking evaporation is not the main process that can induce energy transport. This can also be seen by noting that the Hawking radiation is thermal, i.e. information is not stored in the Hawking radiation. In this sense, some extra processes should be added to transport energy or matter between two sides of the event horizon, so that information will not be lost.

In fact, the Hawing effects are only one part of the total interactions, with the background metric $g^B_{\mu\nu}$ as a classical source. As shown in Eq.~(\ref{d3}), in addition to the Hawking effects, there is also an interaction term between matter and metric perturbation. This interaction term can be formally expressed as
\begin{equation}
\hat{H}_{int}\sim\int dV\hat{h}_{\mu\nu}\hat{T}^{\mu\nu}=\int^{ext} dV\hat{h}_{\mu\nu}[\hat{d}]\hat{T}^{\mu\nu}[\hat{b}]+\int^{int} dV\hat{h}_{\mu\nu}[\hat{\tilde{d}}]\hat{T}^{\mu\nu}[\hat{\tilde{b}}],
\label{g3}
\end{equation}
where the interactions occurring in the exterior and interior regions are separated, and the Killing vector in Eq.~(\ref{d3}) is ignored for simplicity. These two interactions will lead to two unitary evolutions in the two causally disconnected regions. Moreover, they will result in excited states, for example $\hat{b}^{\dag}\arrowvert0_{U}\rangle_{\phi}$ satisfying
\begin{equation}
(\hat{b}^{\dag}\hat{b})\hat{b}^{\dag}\arrowvert0_{U}\rangle_{\phi}\simeq\hat{b}^{\dag}\arrowvert0_{U}\rangle_{\phi}+\gamma(\hat{b}^{\dag})^2\hat{\tilde{b}}^\dag\arrowvert0_{U}\rangle_{\phi},
\label{g4}
\end{equation}
where relation $\hat{b}_{\omega}\arrowvert0_U\rangle_{\phi}=e^{-4\pi M\omega}\hat{\tilde{b}}_{\omega}^{\dag}\arrowvert0_U\rangle_{\phi}$~\cite{s} has been used, with the frequency dependence neglected for simplicity. The in-falling vacuum is used because matter should be able to fall into the black hole interior classically, leading to larger black holes. In other words, the smooth geometry near the event horizon, observed by an in-falling observer, is guaranteed by the existence of the in-falling vacuum. The excited states with creators $\hat{b}^{\dag}$ acted on $\arrowvert0_{U}\rangle_{\phi}$ not only can be treated as the exterior matter, but also can be regarded as the evaporated radiations. Analogously, the superposition of the excited states, such as $\hat{d}^{\dag}\hat{\tilde{d}}^{\dag}\arrowvert0_{U}\rangle_{h}$, can be used to provide the perturbations of the metric in Eq.~(\ref{f2}) by quantum averages.

As long as the entanglements implicit in those in-falling vacua are not destroyed, there will be a (modified) quantum teleportation channel~\cite{n} between the exterior and interior regions, through those interactions in Eq.~(\ref{g3})\footnote{As shown in reference~\cite{n}, the modified quantum teleportation channel can be described by a simplified qubit model. Assume that there are two causally disconnected regions which should be correlated with each other. The correlation can be generated through a EPR pairs, with one qubit in each region. Then by controlling local unitary evolutions in the two regions, the correlation can be established through the entanglement implicit in the EPR pairs.}. The matter or energy can be transported as follows~\cite{n}. To form a black hole with a larger mass, we should add some energy to the matter outside, that is, the scalar field will be at some excited state $(\hat{b}^{\dag})^n\arrowvert0_{U}\rangle_{\phi}$, where the frequency dependence is still ignored for simplicity. The energy stored in the $\hat{b}$ modes can simply be transferred into the $\hat{d}$ modes of the metric perturbation $\hat{h}_{\mu\nu}$, via the ordinary interaction $\hat{b}\hat{b}\hat{d}^\dag+h.c.$ in the exterior region. But how to construct excited states of the form $\hat{\tilde{d}}^{\dag}\arrowvert0_{U}\rangle_{h}$? Note that $\hat{\tilde{d}}^{\dag}\arrowvert0_{U}\rangle_{h}\simeq \lambda\hat{d}\arrowvert0_{U}\rangle_{h}$ due to the entanglement, thus we need some interaction of the form $\hat{b}\hat{b}\hat{d}+h.c.$. This is impossible in a flat space QFT. However, for a black hole background it is possible because \emph{the exterior (or interior) region is not a complete space-time by itself}\footnote{In a (complete) flat space QFT, the interaction $\hat{b}\hat{b}\hat{d}+h.c.$ is forbidden, since the corresponding integral of the mode functions will result in a vanishing delta function $\delta(\omega_1+\omega_2+\omega_3)$. However, in the exterior of the black hole, the integrals are non-vanishing due to the incompleteness of that space-time region.}. After these processes, we will obtain an excited state for the metric perturbation $(\hat{d}^{\dag})^{p}(\hat{\tilde{d}}^\dag)^{q}\arrowvert0_{U}\rangle_{h}$. Certainly, the full state is a superposition of all possible excited states, denoted as $\arrowvert\Psi\rangle_{h,\phi}$ for short, including both the metric perturbation and scalar field. Then the final metric will be $g^B_{\mu\nu}+~_{h,\phi}\langle\Psi\arrowvert\hat{h}_{\mu\nu}\arrowvert\Psi\rangle_{h,\phi}$, and a larger mass black hole may be formed according to Eq.~(\ref{f5}).

Similarly, we can reverse the above process by transporting the energy of a larger mass black hole into the matter outside, leading to a black hole with smaller masses. According to Eq.~(\ref{f5}), the black hole metric with larger mass (together with some small perturbation) can also be represented by another metric with smaller mass together with some larger perturbations. The large perturbations can be expressed as excited states of the metric perturbation, such as $(\hat{d}^{\dag})^p(\hat{\tilde{d}}^{\dag})^q\arrowvert0_{U}\rangle_{h}$ etc.. The energy stored in $\hat{d}$ modes can be transferred into the scalar field's exterior $\hat{b}$ modes still by the ordinary interaction $\hat{b}^\dag\hat{b}^\dag\hat{d}+h.c.$. While the energy stored in the $\hat{\tilde{d}}$ modes can be transferred into $\hat{b}$ modes indirectly as follows. First apply interaction $\hat{\tilde{b}}\hat{\tilde{b}}\hat{\tilde{d}}+h.c.$, then use the relation $\hat{\tilde{b}}\arrowvert0_{U}\rangle_{\phi}\simeq \lambda'\hat{b}^{\dag}\arrowvert0_{U}\rangle_{\phi}$. Certainly, when new black hole has been formed, the mode expansions of those quantum fields will depend on the new black hole's mass. It is this background dependence that makes the complete quantum picture difficult to be constructed. However, except this difficulty, the black hole formation and evaporation can indeed be described in a unitary manner effectively, through the above modified quantum teleportation channel\footnote{The analysis here is only formal, some more detail investigations are still needed.}.

In general, by using of the above quantum channel, matter or energy can go into and come out more easily than that in the classical case. Different from the Hawking effects, information is (almost) not lost since the corrections from the interactions in Eq.~(\ref{g3}) have been included so that all the processes are unitary effectively, except that some background has to be formed in each stage via quantum averages. In order for a black hole to evaporate completely, some extra environment is needed to absorb those evaporated radiations. In reference~\cite{n}, an exterior radiation detector is added to couple with the exterior scalar field modes through ordinary interaction $\hat{b}\hat{e}^\dag+h.c.$, with $\hat{e}^\dag$ the creator of the radiation detector. In this way, the radiations will be represented by the \emph{complete} $\hat{e}$ modes of the exterior radiation detector, instead of the scalar field's \emph{incomplete} $\hat{b}$ modes. Then the energy of evaporated radiations will be given by
\begin{equation}
E_R\simeq~_{\phi,h,D}\langle\Phi\arrowvert\sum_{\omega}\omega\hat{e}_{\omega}^\dag\hat{e}_{\omega}\arrowvert\Phi\rangle_{\phi,h,D},
\label{g5}
\end{equation}
where the state $\arrowvert\Phi\rangle_{\phi,h,D}$ is the full state, including also the interaction between the modes of both the exterior scalar field and the radiation detector. Since the evaporation is unitary effectively except the background dependence, thus the information will not be lost when the black hole evaporates completely, all stored in the remaining components of the closed system, i.e. the scalar field, the radiation detector and the weak gravitational field. In this sense, the information loss paradox is resolved effectively.

Furthermore, the firewall paradox can also be resolved as follows. Through the interaction $\hat{b}\hat{e}^\dag+h.c.$, the radiations have already been represented by the $\hat{e}$ modes of the exterior radiation detector. In this way, the $\hat{b}$ modes are entangled with the radiations in terms of $\hat{e}$ modes. Note further that the radiations in our picture are actually those excited states based on the in-falling vacuum $\arrowvert0_{U}\rangle_{\phi}$, such as $\hat{b}^{\dag}\arrowvert0_{U}\rangle_{\phi}$, as shown below Eq.~(\ref{g4}). While the Hawing radiations are states based on the static vacuum $\arrowvert0,\tilde{0}\rangle_{\phi}$ on the right of Eq.~(\ref{g2}). Thus, our picture of the black hole evaporation is completely different from the Hawking effects. \emph{Our evaporation is based on a quantum channel between the two sides of the event horizon via the entanglement implicit in the in-falling vacuum, not the particle pair production for the Hawking effects based on the static vacuum}. Further, as shown below Eq.~(\ref{g4}), the in-falling vacuum is used to ensure that, the geometry is smooth near the event horizon in the in-falling frame. If the theory was based on the static vacuum, then the in-falling vacuum would decay into the static vacuum via the particle pair production. In this sense, the smooth geometry will be broken or firewall is formed along the event horizon. While in our picture, an in-falling vacuum is always present since the energy transport through those interactions in Eq.~(\ref{g3}) preserves the in-falling vacuum. Therefore, no firewall will be formed since it's not necessary to destroy the entanglements implicit in the in-falling vacua to ensure the monogamy of entanglement. In fact, the $\hat{b}$ modes entangled with the $\hat{e}$ modes are not those entangled with the $\hat{\tilde{b}}$ modes, because they are based on two different vacua, the in-falling vacuum for the former case while the static vacuum for the latter. Thus the monogamy of entanglement is preserved, and the firewall paradox is resolved.

In general, based on a quantum channel between the black hole exterior and interior, the black hole formation and evaporation can be described in a unitary manner effectively. As a result, the black hole thermodynamics may be constructed by constructing statistical mechanics, just like any other ordinary system. This will be shown in the next section.

\section{Black hole thermodynamics revisited }
\label{seci1-3}
In Sec.~\ref{seci1-0}, we have shown that the black hole's first law may not simply be derived by only comparing the mass variation with the thermal first law, due to the deformation term in Eq.~(\ref{d4}). As a consequence, Bekenstein-Hawking entropy may not be Boltzmann or thermal entropy. In this section, we shall give another derivation of black hole thermodynamics that is analogous to the ordinary one. The key point is the underlying (quasi-)stationary point achieved by the equilibrium between the exterior and interior terms in Eq.~(\ref{f4}), at the background of a black hole with some almost fixed mass. And the equilibrium is based on the energy transport through the quantum channel between two sides of the event horizon, as shown in the last section.

To construct thermodynamics, it's necessary to know the internal energy of the studied system. For a \emph{closed} system composed of only black hole and scalar field, the (classical) energy can be given by the total mass in Eq.~(\ref{f4}). In a quantum version, this means that there is an effective Hamiltonian with background dependence\footnote{The background dependence of those individual terms in Eq.~(\ref{h1}) is expressed formally by the black hole's mass $M$ in $\hat{H}_{eff}(M)$ for simplicity.}
\begin{equation}
\hat{H}_{eff}(M)=\hat{H}_0[\hat{b}]+\hat{H}_0[\hat{d}]+\hat{H}_{int}[\hat{b},\hat{d}]+M+\hat{H}_0[\hat{\tilde{b}}]+\hat{H}_0[\hat{\tilde{d}}]+\hat{H}_{int}[\hat{\tilde{b}},\hat{\tilde{d}}],
\label{h1}
\end{equation}
where the black hole's mass $M$ serves as a parameter, which may vary according to the energy transport between the two sides of the event horizon via the quantum channel mentioned in the last section. When the equilibrium between the exterior and interior terms is achieved, the total energy is stable under the equilibrium
\begin{equation}
\delta~_{h,\phi}\langle\Psi\arrowvert\hat{H}_{eff}\arrowvert\Psi\rangle_{h,\phi}\simeq0,
\label{h2}
\end{equation}
where $\arrowvert\Psi\rangle_{h,\phi}$ is the state near the equilibrium point of the closed system. This is one of the conditions for the equilibrium, saying that the energy transport between the two sides of the event horizon is weak enough under the equilibrium. Further, because of the causal barrier of the event horizon, the accessible observable will be the exterior Hamiltonian
\begin{equation}
\hat{H}^{ext}=\hat{H}_0[\hat{b}]+\hat{H}_0[\hat{d}]+\hat{H}_{int}[\hat{b},\hat{d}],
\label{h3}
\end{equation}
with the other terms in Eq.~(\ref{h1}) treated as the observables of some environment. Under this circumstance, another stable condition will be
\begin{equation}
\delta~_{h,\phi}\langle\Psi\arrowvert\hat{H}^{ext}\arrowvert\Psi\rangle_{h,\phi}\simeq0,
\label{h4}
\end{equation}
which means that the energy absorbed and emitted is roughly balanced under the equilibrium. Besides, the black hole's mass is also almost invariant, i.e. $\delta M\simeq0$. And according to Eq.~(\ref{f5}), this corresponds to the following condition
\begin{equation}
~_{h,\phi}\langle\Psi\arrowvert\hat{h}_{\mu\nu}\arrowvert\Psi\rangle_{h,\phi}\ll g^B_{\mu\nu}(M+\Delta M)-g^B_{\mu\nu}(M),
\label{h5}
\end{equation}
which means that the black hole is (quasi-)stationary, without large perturbations or fluctuations that can form black holes with other masses. Except the extra condition for the stationary black hole in Eq.~(\ref{h5}), the other two are familiar conditions in (equilibrium) statistical mechanics.
That is, we obtain a similar picture as the one for the flat space: the exterior studied system and the interior environment, between which a quantum channel can be used to transport energy. While the black hole serves only as some (quasi-)stationary background, guaranteed by the condition in Eq.~(\ref{h5}). In this case, the thermodynamics for the studied exterior system can be obtained by constructing statistical mechanics by means of (grand) canonical ensemble theory.

However, the above construction is only based on the exterior and interior matter and metric perturbations, without the participation of the black hole. Thus it has to extend the above construction by including the black hole in some active sense. To achieve it, the above stable conditions, especially the one in Eq.~(\ref{h5}), should be broken so that all the possible (quasi-)stationary points can occur. If assigning energy levels $\{\epsilon_i(M)\}$ for the respective Hamiltonian operators on the right of Eq.~(\ref{h1}), those (quasi-)stationary points can be denoted effectively as
\begin{equation}
\{M_\alpha,\{\epsilon_i(M_\alpha)\}\},
\label{h5a}
\end{equation}
a semiclassical distribution for the underlying microstates. This may lead to an effective statistical mechanics by ignoring the details within each black hole's mass $M_\alpha$. For example, the partition function can be obtained in the continuous limit of the black hole's mass\footnote{The trace in Eq.~(\ref{h5b}) should be operated in terms of states based on the in-falling vacua at some specific background given by some mass $M$, as analyzed in Sec.~\ref{seci1-2}.}
\begin{equation}
\mathcal{Z}_{eff}(\beta)=\int_{0}^{M_{max}} dM~tr~e^{-\beta\hat{H}_{eff}(M)}=\int_{0}^{M_{max}} dM\sum_i\exp\{-\beta[M+\epsilon_i(M)]\},
\label{h5b}
\end{equation}
with $M_{max}$ the maximum black hole's mass for a closed black hole-matter system\footnote{The closed black hole-matter system should be described by a microcanonical ensemble, here a canonical ensemble is used for simplicity.}. While for an open black hole-matter system coupling with an exterior heat bath\footnote{The interaction between the black hole-matter system and the heat bath can be given by $\hat{b}\hat{e}^\dag+h.c.$, just like the case for the radiation detector introduced in the last subsection.}, the upper limit of integral can be infinity. Unfortunately, the partition function in Eq.~(\ref{h5b}) is hard to calculate because of the complicated background dependence in the effective Hamiltonian.

Then how many microstates are stored in a black hole, or what is the Boltzmann entropy of a black hole with some given mass? Obviously, the Bekenstein-Hawking entropy cannot be the correct answer, according to the previous discussions in Sec.~\ref{seci1-0}. Actually, as indicated by Eq.~(\ref{f5}), the black hole can be formed step by step by some quantum averages of the quantum gravitational fields in each stage, as shown in Sec.~\ref{seci1-2}. In the view of this effective field picture, the microstates within a black hole should be provided by those averaged quantum states which cannot be counted easily. Therefore, a more complete quantum gravity theory is still needed to answer the question. However, we may give a coarse answer as follows. The expression in Eq.~(\ref{f5}) can be roughly extrapolated to an extreme case
\begin{equation}
g^B_{\mu\nu}(M)+h'_{\mu\nu}=\eta_{\mu\nu}+h^m_{\mu\nu},
\label{h6}
\end{equation}
that is, a black hole metric can be represented by a flat space metric with large perturbations caused by concentrating large amount of matter within some finite space. Although the process is hard to describe in a quantum manner, it should be unitary in general. This indicates that the effective Hamiltonian in Eq.~(\ref{h1}) can also be extrapolated to the corresponding extreme case
\begin{equation}
\hat{H}(\eta)=\hat{H}_0[\hat{a}]+\hat{H}_0[\hat{c}]+\hat{H}_{int}[\hat{a},\hat{c}],
\label{h7}
\end{equation}
i.e. an ordinary flat space Hamiltonian in terms of modes that can be identified with those in-falling modes approximately. Thus the formation and evaporation of a black hole can be expressed formally by a sequence of Hamiltonians
\begin{equation}
\hat{H}(\eta)\rightarrow\hat{H}_{eff}(M_1)\rightarrow\hat{H}_{eff}(M_2)\rightarrow\cdots\rightarrow\hat{H}(\eta),
\label{h8}
\end{equation}
giving an effective unitary evolution except the background dependence. When the black hole's mass satisfies $M\simeq\langle\hat{H}(\eta)\rangle$, then the entire system's energy is almost stored in the black hole. In other words, almost all the matter has fallen into the black hole, except some tiny quantum fluctuations. Under this circumstance, \emph{the Boltzmann entropy for the final black hole will roughly be identified with that of the in-falling matter together with the initial gravitational field}. This is mainly because all the evolutions are unitary without information loss, and the initial information should be almost stored in the final black hole\footnote{This fact indicates further that the Bekenstein-Hawking entropy, which is roughly the event horizon's area, may not be a Boltzmann entropy. Otherwise, information will be lost inevitably.}.
Moreover, given an initial Hamiltonian in Eq.~(\ref{h7}), the black hole-matter system's thermodynamics can also be constructed by constructing statistical mechanics. For instance, the partition function can simply given by $Tr e^{-\beta\hat{H}(\eta)}$ formally.

In general, the black hole thermodynamics can be derived ordinarily by constructing statistical mechanics, provided that the black hole formation and evaporation can be described in a unitary manner. Certainly, the background dependence make the construction more difficult than the ordinary system in a flat space. A complete quantum gravity theory without background dependence is still needed.

\section{Summary and conclusions}
\label{secv}

In this paper, black hole thermodynamics is constructed according to the unitary evolutions of the black hole-matter system. We show that the Bekenstein-Hawking entropy $S_{BH}$ may not be treated as a Boltzmann entropy. This is clarified by a reconsideration of the original derivation of the black hole's ``first law", in which metric perturbations caused by matter were ignored. After including those metric perturbations, the formula of total mass variation for the black hole-matter system cannot simply be treated as the first law of thermodynamics formally, implying that $S_{BH}$ may not be a Boltzmann entropy. Further, by including those (quantum) metric perturbations, the event horizon can be tunneled effectively through a quantum channel, so that energy can be absorbed and emitted more efficiently. As a result, the black hole formation and evaporation can be described in a unitary manner effectively. Then, based on the unitary evolutions, the black hole thermodynamics can be constructed by first constructing statistical mechanics.

It thus seems that black hole thermodynamics \emph{may not} be well described in terms of Bekenstein-Hawking entropy $S_{BH}$ and Hawking temperature $T_{H}$. This is mainly because the black hole thermodynamics in terms of $S_{BH}$ and $T_{H}$ are not based on the system's unitary evolutions.
As a consequence, there is no thermal character for pure gravity, only from the view of Einstein's equation~\cite{u,v}. Moreover, the gravity is still a fundamental force, no entropy force~\cite{w} is necessary. Certainly, the analysis in this paper is only semiclassical, and there still needs a full quantum gravity theory to describe the details of the interactions between gravity and matter.

\acknowledgments

This work is supported by the NSF of China, Grant No. 11375150.


\begin{thebibliography}{99}

\bibitem{a} S.W. Hawking, \emph{Particle creation by black holes}, \emph{Commun. Math. Phys.} {\bf 43}, (1975) 199.


\bibitem{b} S.W. Hawking, \emph{Breakdown of predictability in gravitational collapse}, \emph{Phys.Rev.} {\bf D 14}, (1976) 2460.

\bibitem{c} J.M. Bardeen, B. Carter and S.W. Hawking, \emph{The Four Laws of Black Hole Mechanics}, Comm. Math.
Phys. {\bf 31}, (1973) 161.

\bibitem{d} J.D. Bekenstein, \emph{Black Holes and Entropy}, \emph{Phys. Rev} {\bf D 7}, (1973) 2333.

\bibitem{e} J. Preskill, \emph{Physics 229: Advanced Mathematical Methods of Physics, Quantum Computation and Information}, California Institute of Technology (1998).

\bibitem{f} Michael A. Nielsen and Isaac L. Chuang, \emph{Quantum Computation and Quantum Information}, Cambridge University Press (2010).


\bibitem{g} R.D. Sorkin, in \emph{General Relativity and Gravitation, proceedings of the GR10 Conference}, Padova, 1983,
edited by B. Bertotti, F. de Felice, and A. Pascolini (Consiglio Nazionale delle Ricerche, Roma, 1983), Vol.2.

\bibitem{h} L. Bombelli, R.K. Koul, J. Lee and R.D. Sorkin, \emph{A quantum source of entropy for black holes}, \emph{Phys. Rev} {\bf D 34}, (1986) 373.

\bibitem{i} V. Frolov and I. Novikov, \emph{Dynamical origin of the entropy of a black hole}, \emph{Phys. Rev} {\bf D 48}, (1993) 4545.

\bibitem{i1} Mark, Srednicki, \emph{Entropy and area}, \emph{Phys. Rev. Lett} {\bf 71}, (1993) 666.

\bibitem{j} L. Susskind and J. Uglum, \emph{Black hole entropy in canonical quantum gravity and superstring theory}, \emph{Phys. Rev} {\bf D 50}, (1994) 2700.

\bibitem{j1} Sergey Solodukhin, \emph{Entanglement entropy of black holes}, arXiv:1104.3712v1 [hep-th].

\bibitem{k} A. Strominger and C. Vafa, \emph{Microscopic origin of the Bekenstein-Hawking entropy}, \emph{Phys. Lett} {\bf B 379}, (1996) 99.

\bibitem{l} G. 't Hooft, \emph{Dimensional reduction in quantum gravity theory}, gr-qc/9310006 (1993).

\bibitem{m} L. Susskind, \emph{The world as a hologram}, \emph{J. Math. Phys} {\bf 36}, (1995) 6377.

\bibitem{n} Yu-Lei Feng and Yi-Xin Chen, \emph{A Unitary Model of The Black Hole Evaporation}, \emph{JHEP} {\bf 12} (2014) 088.

\bibitem{n1} A. Almheiri, D. Marolf, J. Polchinski, and J. Sully, \emph{Black Holes: Complementarity or Firewalls}?, arXiv:1207.3123 [hep-th].

\bibitem{n2} A. Almheiri, D. Marolf, J. Polchinski, D. Stanford, and J. Sully, \emph{An Apologia for Firewalls}, arXiv:1304.6483 [hep-th].

\bibitem{n3} S. L. Braunstein, S. Pirandola and K. Zyczkowski, \emph{Better Late than Never: Information Retrieval from Black Holes}, \emph{Phys.Rev.Lett} {\bf 110}, (2013) 101301.

\bibitem{n3a} Steven B. Giddings, \emph{Statistical physics of black holes as quantum-mechanical systems}, \emph{Phys. Rev} {\bf D 88}, (2013) 104013.

\bibitem{n4} Ted Jacobson, \emph{Introductory Lectures on Black Hole Thermodynamics}, 1996.

\bibitem{s} Alessandro Fabbri and Jos¨¦ Navarro-Salas, \emph{Modeling Black Hole Evaporation}, World Scientific Publishing Company (2005).

\bibitem{u} Ted Jacobson, \emph{Thermodynamics of Spacetime: The Einstein Equation of State}, \emph{Phys.Rev.Lett} {\bf 75}, (1995) 1260.

\bibitem{v} T. Padmanabhan, \emph{Thermodynamical aspects of gravity: new insights}, \emph{Rep. Prog. Phys} {\bf 73} (2010) 046901.

\bibitem{w} Erik Verlinde, \emph{On the origin of gravity and the laws of Newton}, \emph{JHEP} {\bf 04} (2011) 029.

\end{thebibliography}
\end{document}